\documentclass[apj]{emulateapj}
\usepackage{apjfonts}
\usepackage{natbib}
\usepackage{psfig}
\usepackage{amsmath}
\def\be{\begin{eqnarray}}
\def\ee{\end{eqnarray}}
\usepackage{comment}
\usepackage{graphicx,subfigure}
\usepackage{appendix}
\usepackage{enumerate}

\shorttitle{Interstellar Time Delay correction}
\shortauthors{Palliyaguru, Stinebring, McLaughlin, Demorest \& Jones}

\begin{document}
\title{Correcting for Interstellar Scattering Delay in High-precision Pulsar Timing: Simulation Results}
\author{Nipuni Palliyaguru\altaffilmark{1}, Daniel Stinebring\altaffilmark{2,3}, Maura McLaughlin\altaffilmark{1,4}, Paul Demorest\altaffilmark{5} \& Glenn Jones\altaffilmark{6}}
\affil{
$^{1}$ Department of Physics, West Virginia University, Morgantown, WV 26506, USA\\
$^{2}$ Department of Physics and Astronomy,110 North Professor Street, Oberlin College, Oberlin, OH 44074, USA\\
$^{3}$ ASTRON, The Netherlands Institute for Radio Astronomy, Postbus 2, 7990 AA, Dwingeloo, The Netherlands\\
$^{4}$ Adjunct Astronomer, the National Radio Astronomy Observatory, Charlottesville, VA 22903, USA\\
$^{5}$ National Radio Astronomy Observatory, 520 Edgemont Road, Charlottesville, VA 22903, USA\\
$^{6}$ Department of Physics, Columbia University, New York, NY, 10027, USA\\
npalliya@mix.wvu.edu, dan.stinebring@oberlin.edu, maura.mclaughlin@mail.wvu.edu, pdemores@nrao.edu, glenn.caltech@gmail.com
}

\begin{abstract}
Light travel time changes due to gravitational waves may be detected within the next decade through precision timing of millisecond pulsars.
Removal of frequency-dependent interstellar medium (ISM) delays due to dispersion and scattering is a key issue in the detection process.
Current timing algorithms routinely correct pulse
times of arrival (TOAs)  for time-variable delays due to cold plasma dispersion. However, none of the major pulsar timing groups correct for delays due to scattering from multi-path propagation in the ISM.
%, which has a more complicated scaling with frequency, but generally is considered to have a term scaling as $\nu^{-x}$, with $x \approx 4.0 - 4.4$.
Scattering introduces a frequency-dependent phase change in the signal that results in pulse broadening and arrival time delays. 
Any method to correct the TOA for interstellar propagation effects must be based
on multi-frequency measurements that can effectively separate dispersion and scattering delay terms from frequency-independent perturbations such as those due to a gravitational wave.
Cyclic spectroscopy, first described in an astronomical context by \citet{d11}, is a potentially powerful tool to assist in this multi-frequency decomposition. 
As a step toward a more comprehensive ISM propagation delay correction, we demonstrate through a simulation that we can accurately recover impulse response functions (IRFs), such as those that would be introduced by multi-path scattering, with a realistic signal-to-noise 
ratio. We demonstrate that timing precision is improved when scatter-corrected TOAs are used, under the assumptions of a high signal-to-noise and highly scattered signal. 
%We also show that, for a simple test case, we can isolate these delays from other types of delays.
%Reductions in the timing residual root-mean-square of more than a factor of two are possible through removal of time-variable scattering delays. 
%Additional development that makes use of the frequency dependence of scattering is likely necessary for a realistic implementation in pulsar timing array data.
%with consequent improvements in timing precision. 
%We also demonstrate that we can isolate the scattering delays from other types of delays, and show that reductions in the timing residual root-mean-square of more than a factor of two are possible through removal of time-variable scattering delays.
We also show that the effect of pulse-to-pulse ``jitter'' is not a serious problem for IRF
reconstruction, at least for jitter levels comparable to those observed in several
bright pulsars.

\end{abstract}


\maketitle

\section{Introduction}

Gravitational waves (GWs) are a key prediction of Einstein's theory of general relativity 
and their existence has been supported through timing measurements of the orbital decay of the Hulse-Taylor 
binary system B1913$+$16 \citep{ht75}. Many experiments aim to detect these waves directly through the measurement of light travel changes between objects.
Complementary to interferometer-based GW detection experiments
like LIGO, pulsar timing is sensitive to nanohertz frequency
GWs. The change in light travel time between the earth
and a pulsar due to a passing GW results in a delay in the time of arrival (TOA) of pulses \citep{d79}. 
Given a timing model that accounts for parameters such as pulsar period, period derivative, position, proper motion and other orbital parameters, we calculate residuals, or the differences between measured and model TOAs.   

These residuals will contain the signatures of gravitational waves. 
A stochastic background of GWs can be detected through searching for a correlation with angular separation in the timing residuals of an array of pulsars \citep{hd83}.
In order to detect the background due to supermassive black hole binaries, over 40 MSPs with root-mean-square (RMS) timing residuals
of less than 100 ns are likely required \citep[][]{jhl05,cs11}. 
Currently over 40 millisecond pulsars are being timed by the North
American Nanohertz Observatory for Gravitational waves (NANOGrav), with
RMS timing residuals of nearly all pulsars at the sub-microsecond level
\citep{dfg12,mclaughlin13}. In addition to GWs, other
effects such as interstellar medium (ISM) propagation and
rotational irregularities will affect the arrival times of pulses.
Fortunately, ISM effects are chromatic and therefore  multi-frequency
observations can be used to at least partially correct for these variations.

Electromagnetic radiation from pulsars experiences delays as it travels through the ionized plasma of the ISM.
The three prominent known effects  are (1) dispersion, caused by the change in radio wave speed due to refraction, (2) scattering and scintillation, due to inhomogeneities in the medium \citep{r69} which results in a random interference pattern on the observer plane, and (3) Faraday rotation, which is rotation of the plane of linear polarization due to a magnetized plasma.
All timing algorithms correct for time-variable dispersion to high accuracies \citep {kcs13}.
We do not expect Faraday rotation to result in TOA fluctuations if polarization calibration is done
correctly. 
In this paper we concentrate on removal of scattering effects, which are more difficult to correct but can cause sizeable fluctuations in TOAs.

\begin{figure*}
    \centering
\includegraphics[width=3.3in]{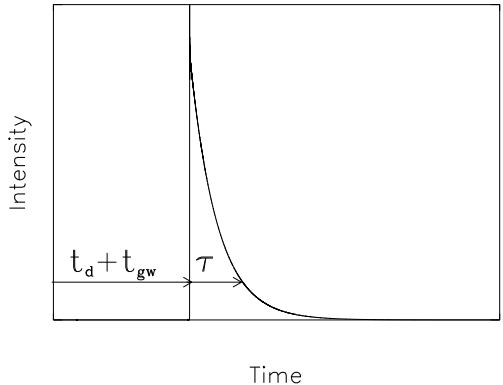}
\caption{A simplified model that describes how some of the most prominent effects cause delays in the pulsar signal. Delays from gravitational waves, dispersion, and scattering of the intensity IRF are given by $\rm t_{gw}$, $\rm t_{d}(\nu)$ and $\tau(\nu)$, respectively.
The frequency dependence of dispersion and scattering delays scale as $\nu^{-2}$ and $\sim\nu^{4}$ respectively.
The former two effects cause time delays of the pulse while scattering also changes its shape, broadening it and thereby causing an additional delay.
While the dispersive delay $\rm t_{d}(\nu)$ also broadens the pulse, it is usually corrected for in pulsar data processing through coherent de-dispersion, for any line-of-sight that has a unique DM.}
\label{fig:toymodel}
\end{figure*}

The distribution of electron density in the ISM can be described by the spatial spectrum of turbulence, or the spectral density, 
\begin{equation}
P_{ne}\left(q\right)=\frac{C_{ne}^2}{\left(q^2+\kappa_0^2\right)^{\beta/2}}\rm exp\left(-\frac{q^2}{4\kappa_i^2}\right), 
\end{equation}
where $q$ is the wave number, $\beta$ is the spectral exponent and $\kappa_i^{-1}$ and $\kappa_0^{-1}$ are the inner and outer scales respectively \citep{r90}.
For $\kappa_0\ll q \ll \kappa_i$ it is approximated by a power law $q^{-\beta}$, where
$\beta=11/3$ for a Kolmogorov medium, which describes a cascade of kinetic energy in the interstellar plasma \citep{lr99}.
This generally describes most pulsar lines of sight \citep{gr94}, even though inconsistencies may exist for others \citep{kcs13}.
The phase structure function, or the mean square phase difference between neighboring ray paths, 
takes the form of a power law for $\beta<4$.
For steeper spectra, the phase structure function will be a square law
\citep{ar95,lr00}.
The scattering process delays the pulse TOA due to refraction and multipath propagation.
While the most prominent scattering effect is pulse broadening due to multipath propagation,  
other effects such as angle of arrival variations also contribute to the pulse delay.
Scintillation causes the pulse to appear brighter at certain times and frequencies, with characteristic scales determined by the distance to the pulsar, its
velocity,  the properties of the ISM along the line of sight, and the observing frequency;
for a review of these effects see \citet{s13}.
Figure~\ref{fig:toymodel} shows a schematic picture of how these various delays affect the signal.
The long-term goal of pulsar timing is to correct for delays due to other intrinsic and extrinsic effects such that only the GW signature remains in the residuals.

ISM delays are observing frequency ($\nu$) dependent, with dispersion and pulse broadening scaling as $\nu^{-2}$ and $\nu^{x}$, respectively, with $x\approx-4$ \citep{lk05}. Therefore pulse broadening, which is indicative of large amounts of scattering, is most prominent at low frequencies.
In addition to the frequency dependence, pulse broadening has been empirically determined to have a roughly DM$^{2}$ dependence  \citep{bcc04}, where DM is the dispersion measure, or the integral of the electron density along the line of sight. 
Scattering delays are also expected to vary significantly with time due to the relative motion of the pulsar and the Earth changing the line-of-sight path through the ionized ISM.
\citet{hs08} used secondary spectra of pulsar B1737$+$13 to measure scattering delays between 0.2 and 2.2 $\mu$s over $\sim$270 days of observation at a radio frequency of 1400 MHz.
\citet{rdb06} showed that scattering delays vary between $\sim100$ and $\sim140$ us over $\sim10$ years for B1937$+$21 at 327 MHz.

Correcting for ISM scattering delays may be important for detecting GW signatures in our data \citep[e.g.,][]{fc90}.
In addition, the average spectral index of millisecond pulsars is $\sim$--1.4 \citep{blv13}, meaning that these objects are $\sim$8 times brighter at 430 MHz than at  1400 MHz.
The MSPs used in current timing experiments are selected to be nearby (i.e. $<$few kpc) and are generally timed at high frequencies ($\geq$800 MHz) in order to mitigate these dispersion and scattering effects.
The ability to correct for the effects of scattering could improve timing at lower frequencies, resulting in increased signal-to-noise ratio (S/N).
Finally, understanding the scattering phenomenon will lead to better quantifying the Galactic models for free electron density \citep{cl02} and the distribution of scattering material along the line of sight \citep{cr98}.

Several methods have been proposed to estimate scattering timescales.
These methods assume that the ISM acts as a linear filter with a voltage impulse response function (IRF), which is convolved with the intrinsic pulsar signal to produce the observed pulse.
For scattering by a single thin screen with a square law structure function \citep{cr98}, the ensemble-average IRF is a one-sided exponential.
\citet{ki93} showed that a descattered pulse can be restored by fitting the observed profile to a Gaussian convolved with a one-sided exponential function.
However, these methods usually require assumptions about the functional form of the IRF, which is dependent on the spatial distribution and inhomogeneity spectrum of the scattering medium \citep{cr98}.
The scattering times can also be estimated from the auto correlation function (ACF) of the pulsar dynamic spectra or from the cumulative delay function from pulsar secondary spectra \citep[see, e.g.,][]{hs08}.
However, these methods are limited by large uncertainties and a finite number of scintles within the observing bandwidth and observation time.
Another method, which is based on a CLEAN algorithm, tests various IRF types to get the best fit, and requires assumptions about the IRF form \citep{bcc03}. This method can estimate IRFs when scattering delays are large and cause recognizable changes to pulse shapes, but is not optimal in the case of small delays.
More recently \citet{crg10} showed that scattering is anti-correlated with pulse power and the TOA fluctuations can be reduced by $\sim$25\% by removing those correlated components. 

Unlike these methods, cyclic spectroscopy (CS) directly accounts for phase changes of the electric field,  thereby allowing a more accurate description of ISM effects.
The phase of the ISM transfer function, which is the frequency-domain representation of the IRF, contains information about pulse broadening due to ISM delays.
Recovering the phase information of the electric field to reconstruct the IRF has been successfully applied to pulsar dynamic spectra \citep{wkd08}. 
In this paper we explore the deconvolution technique of CS, introduced in \citet{d11} and further developed in \citet[][hereafter WDS13]{wdv13}.
This method allows determination of the phase of a periodic signal, which can then be used to calculate scattering delays.
This paper is a step in an ongoing analysis of the efficiency of CS for scattering delay correction.
By means of a simulation that includes a realistic signal model, we show that CS can be used to 
accurately
reconstruct the IRF for an achievable  signal-to-noise ratio.
We introduce the theoretical formulation in Section~2; in Section~3 we present  the details and results of our simulation; and
in Section~4 we discuss future applications and the advantages of using CS over other methods.

\section{Theoretical background}

We  begin by expressing
the electric field vector as a function of time $t$ and position $\mathbf{r}$, as $\mathbf{E}(\mathbf{r},t)=\mathbf{E_{0}}{\rm exp}\left[-i(\it{\mathbf{k}\cdot \mathbf{r}}-\omega t)\right]$, where $\mathbf{k}$ and $\omega$ are the wave vector and angular frequency of the wave respectively.
For a wave with frequency $f$, traveling in a medium with refractive index $\mu$ at a speed $c$, the wave vector is $k=(2\pi/c)\mu f$.
Frequency-dependent refractive index fluctuations in the medium cause a change in $k$, which corresponds to a change in the phase of the wave.
Upon encountering a scattering region, in this case a thin screen of thickness $a$,
the phase of the wave changes by an amount $\Delta\Phi = \Delta k a$ \citep[see, e.g.,][for details]{lk05}.
For a wave propagating in the $\mathbf{\hat z}$ direction, the phase of the wavefront becomes a function of x and y after passing through the phase screen.
The phase changes will vary randomly along the wavefront, and hence the final phase changes will be randomly distributed.
The final phase `corrugation', causes angular broadening of the propagating radiation.
Additionally, electron density variations that are large compared to the broadened ``image'' of the pulsar projected on the scattering screen will result in refractive effects. See \citet{r90} for a review. 
The bent wavefront arrives later than the unscattered one, resulting in a scattering response that approximates an  exponentially-decaying function in long-term averages \citep{w72}.

In general, pulsar (E-field) signals can be considered to be amplitude-modulated complex Gaussian noise \citep{r75} so that
\begin{equation}
x(t)=n_1(t)p(t), 
\label{eq:amnprof}
\end{equation}
where $\it p(t)$ is a real-valued, positive definite pulse modulation function and $n_1(t)$ is complex Gaussian white noise.

The interstellar medium propagation can be modeled as a linear filter with a voltage IRF
%Under the assumption that the interstellar medium affects only the phase of the propagating signal and not its amplitude, it acts as a linear filter with a voltage BF 
$\it h(t)$ and a corresponding transfer function $\it H(\nu)$, where $\it h(t)$ and $\it H(\nu)$ are a Fourier transform pair. 
For an intrinsic pulsar voltage (E-field) signal $\it x(t)$, the observed voltage signal in the time domain will be $\it y(t)=x(t)*h(t)$, where * denotes a convolution. This voltage signal $y(t)$ plus additive noise, discussed further below, is recorded by baseband observing systems.
The frequency-domain representation of the signal will be  $\it Y(\nu)=X(\nu)H(\nu)$, where $\it x(t)$ and $\it X(\nu)$ are a Fourier transform pair.

Following the notation of \citet{d11}, the ``cyclic spectrum'' as a function of radio frequency $\nu$ and cycle frequency $\alpha$ is then given by \citep{a07,g91} 
\begin{equation}
S_{y}\left(\nu;\alpha\right) = E\left\{Y\left(\nu+\frac{\alpha}{2}\right)Y^{*}\left(\nu-\frac{\alpha}{2}\right)\right\},
\label{eq:csfromprof}
\end{equation}
where $Y^{*}(\nu)$ is the complex conjugate of $Y(\nu)$ and $E$ represents the expectation value.
For a true cyclostationary signal to have non-zero $S_{y}(\nu;\alpha)$, the cyclic frequency $\alpha$ must take on discrete values such that $\alpha_n=n/P$, where $P$ is the pulse period. 
The cyclic spectrum can be further expanded as
%\begin{equation}
\begin{multline}
S_{y}\left(\nu;\alpha_n\right)=E\left\{X\left(\nu+\alpha_{n}/2\right)X^{*}\left(\nu-\alpha_{n}/2\right)\right\}\times\\ 
H\left(\nu+\alpha_{n}/2\right)H^{*}\left(\nu-\alpha_{n}/2\right).
\end{multline}
\label{eq:eq3}
%\end{equation} 
Assuming that the pulsar flux within the band is $S_0$ and $c_{n}$ is the $n^{th}$ complex Fourier coefficient of the Fourier transform of the intensity modulation function $p_{I}(t)$, where $p_{I}(t)=p(t)^2$, we can express the spectrum of the intrinsic signal as
\begin{equation}
E\left\{X\left(\nu+\alpha_{n}/2\right)X^{*}\left(\nu-\alpha_{n}/2\right)\right\}=c_{n}S_0.
\label{eq:eq4}
\end{equation}
Therefore, the cyclic spectrum reduces to
%\begin{multline}
\begin{equation}
S_{y}\left(\nu;\alpha_n\right)=c_n S_0 H\left(\nu+\alpha_{n}/2\right)H^{*}\left(\nu-\alpha_{n}/2\right).
%\end{multline}
\label{eq:eq5} 
\end{equation}
The deconvolution algorithm in \citet{d11} and WDS13 models the cyclic spectrum with an initial input of a delta function transfer function $\it H(\nu)$ and iterative fitting to arrive at the recovered IRF.
We use the publicly available Python version\footnote{https://github.com/gitj/pycyc} of this algorithm.
The IRFs are calculated from the cyclic spectra via a least square minimization of the difference between the modeled and actual cyclic spectrum.
The initial guess for the intrinsic profile used in the analysis was initialized from the data. 
%you give it an initial guess and then it finds the best H(f). 
The intrinsic profile is then extracted given the data and the recovered $H(\nu)$.
In more sophisticated applications, many iterations can be used to refine the template.

\section{Simulations}

In this Section we use simulated data to test the effectiveness of the deconvolution algorithm in recovering IRFs and the effect of scatter correction for pulsar timing.
We first consider scattering by a thin screen and present the effect of scatter correction in Subsection 3.1 and the result of including non-scattering delays such as GWs in Subsection 3.2. 
Then we consider the effect of scatter correction when the ISM is described by a thick screen in Subsection 3.3, and effect of pulse-to-pulse jitter in Subsection 3.4.
We start by forming a pulsar signal $x(t)=n_1(t)p(t)$, as outlined in Equation~\ref{eq:amnprof}.
For simplicity, we choose $p(t)={\rm exp}\left[-(t/W)^2\right]$, a Gaussian-shaped modulation function whose width is $W$, and where $t$ spans the pulse period for a single pulse, and repeats itself to infinity.
Strictly speaking, this equation describes only a single pulse.
Variations in the single pulses are caused by varying $n_1(t)$ values.

For a scattering medium approximated by a single thin screen, and assuming that the refractive index fluctuations within the screen have a square-law structure function \citep[e.g.,][]{cr98}, the voltage IRF has a one-sided exponential envelope \citep{c76} and takes the form
\begin{equation}
h(t)=n_2(t)U\left(t-t_{0}\right) \,{\rm exp}\left[-\frac{t-t_{0}}{\tau^{\prime}}\right], 
\label{eq:ht}
\end{equation}
where $n_2(t)$ is complex Gaussian white noise, $U(t)$ is the unit step function, $t_{0}$ is any uncorrected time delay (relative to a fiducial pulse template), and $\tau^{\prime}=2\tau$, where $\tau$ is the characteristic width in the intensity IRF, which will be referred to as the scattering timescale hereafter.
The inclusion of $n_2(t)$ in Equation~\ref{eq:ht} allows each run of the simulation to be an example of a ``snapshot image" of the ISM \citep{ng89} for each realization of $n_2(t)$ and incorporates the effect of scintillation.
We form the Fourier transform of the observed voltage signal $y(t)$ using the above transfer function 
\begin{equation}
Y(\nu)=X(\nu)H(\nu)+N_{sys}(\nu)
\label{eq:single_prof_freqdom}
\end{equation} 
where $N_{sys}$ is complex additive instrumental noise.

We simulate a pulsar with a period $P$, a pulse width $W$, where $P\gg W$, and a scattering timescale $\tau$, where $\tau<W$.
The width of a pulse $W_{a}$, composed of a noisy Gaussian with width $W_{i}$, convolved with a one-sided-exponential voltage IRF having a broadening timescale $\tau$, can be expressed as $W_a\approx\sqrt[]{W_{i}^2+\tau^2}$.
As the scattering timescale $\tau$ increases and becomes comparable to the pulse width, shape changes come into play, in addition to time delays when cross-correlating the standard and observed profiles.
We do not consider this case here.
In particular, we consider a period of 1.6 ms, a pulse width of 40 $\mu$s in the intensity profile, and a mean scattering timescale of 5 $\mu$s in the intensity IRF $\it h_{I}$, where $h_{I}=|h(t)|^{2}$.
These quantities are similar to the values for the bright MSP, B1937+21, observed at a frequency of 1 GHz.
Amplitude modulated noise was produced in the frequency-domain as Gaussian white noise, with the middle half of the spectrum removed, to allow for oversampling by a factor of two. The noise was then Fourier-transformed so it would be correlated in the time domain.
 This was then multiplied with the Gaussian pulse modulation function $p(t)={\rm exp}\left[-(t/W)^2\right]$, as indicated in Equation~\ref{eq:amnprof}, to produce a single pulse as emitted at the pulsar.
The frequency-domain representation of the observed single pulse waveform is then calculated using Equation~\ref{eq:single_prof_freqdom}, with the simulated ISM transfer function $H(\nu)$, which is the Fourier transform of $\it h(t)$ as given in Equation~\ref{eq:ht}.

We simulate the signal path as closely as possible, including the production of individual pulses, in order to ensure that subtle reconstruction effects are not overlooked.
These simulated single pulses are used to compute the cyclic spectra via the frequency-domain approach, as outlined in Equation~\ref{eq:csfromprof}. 
We have added Gaussian white noise to the cyclic spectrum in the frequency domain to simulate instrumental noise. 
The pulse profiles and cyclic spectra were obtained from averaging $N_p=10^{4}$ single pulses, using a bandwidth of 5 MHz.

The amplitude and phase of simulated cyclic spectra as functions of frequency are shown in Figure~2.
Figure~\ref{fig:PBF_WDS} shows the reconstructed intensity IRF.

\begin{figure*}
\centering
{
\includegraphics[angle=0,width=3.3in]{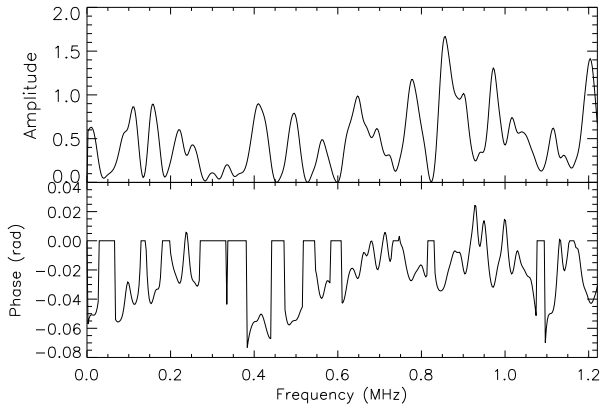}   
\label{fig:cs}
    }
\caption{The amplitude (top) and phase (bottom) of a cyclic spectrum corresponding to a pulse profile with a  period of 1.6 ms and a pulse width of 40 $\mu$s, as a function of frequency, for the first cycle frequency ($n=1$). 
The phase has been set to zero, only for plotting purposes, for very small amplitudes in the cyclic spectrum.
The noise is self-noise from the pulsar amplitude-modulated noise process after an integration of $N=10^{4}$ pulses.
Instrumental noise was not included in this simulation.
}
\end{figure*}

\begin{figure}[h]
\begin{center}
\includegraphics[width=3.5in]{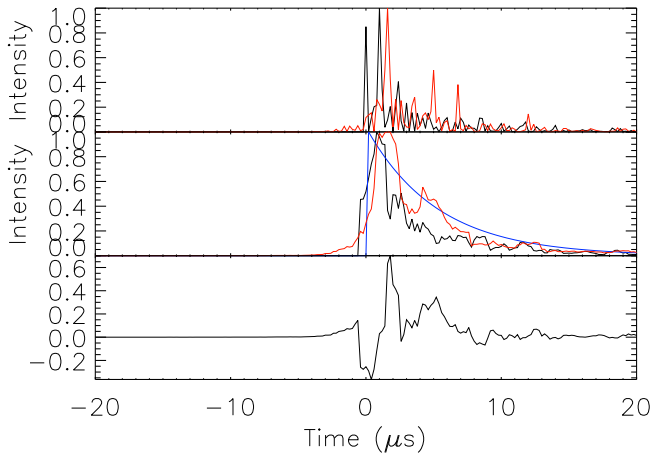}
\label{f1}
\caption{An example recovered normalized intensity IRF $h_{I}(t)$ constructed from a cyclic spectrum (red curve)
for an input scattering timescale of 5.1 $\mu$s. For this realization, the scattering timescale estimated using the fitting technique is 5.7 $\mu$s.
The input function is shown by the black curve and the best fitting one-sided exponential function is shown by the blue curve. 
The middle plots are smoothed versions of the top plots and 
the bottom plot shows the residual between the input and recovered IRFs.}
\label{fig:PBF_WDS}
\end{center}
\end{figure}

\subsection{Scatter correction}

The scattering process broadens the pulse and delays the TOA by shifting its centroid \citep[see, e.g.,][]{crg10}.
In practice, there will be other chromatic and achromatic delays in the data, such as those due to refraction, parallax, proper motion and GW signals, in addition to scattering, which will give a non-zero $t_0$ value in the IRF outlined in Equation~\ref{eq:ht}.
%iTherefore, $t_0$ can be considered as the scatter-corrected TOA. %, with all interesting time signatures remaining intact. 
Therefore, we can approximate the non-scattered TOA, with all interesting time signatures intact, as $t_0$.
%In order to calculate $t_0$, we fit a one-sided exponential function to the recovered PBF in order to locate the rising edge of the PBF and use these $t_0$ values as our scatter-corrected TOAs.
We list the key processing steps in gauging the effectiveness of CS and scatter correction here and describe them in detail in the subsequent paragraphs.
\begin{enumerate}
\item Calculate the TOAs of the scatter-broadened profiles using the Taylor algorithm and a scatter-broadened template.
\item Calculate timing residuals, assuming the constant pulse period used to create the profiles.
\item Reconstruct the IRF from the cyclic spectra using the WDS13 algorithm.
\item Fit a one-sided exponential function to the recovered IRF to return $t_0$, the rising edge of the IRF, or the scatter-corrected TOA.
\item Compare the RMS residuals of the TOAs from scattered profiles and the scatter-corrected TOAs.
\item Repeat this process for varying profile S/N to test how the RMS ratio between scatter corrected and scattered residuals varies with profile S/N.
\item Repeat this process for varying mean scattering timescales to test how the RMS ratio between scatter corrected and scattered residuals varies with mean scattering timescale.
\end{enumerate}

We simulated average pulse profiles and cyclic spectra for 25 trials, with each trial representing a single epoch, with time-variable scattering, where the scattering timescale $\tau$ was drawn from a random distribution with a variance 1 $\mu$s. We have only considered cases in which  $\tau<W$.
The scattering timescales of input one-sided exponential intensity IRFs ranged from $\sim$3.3 $\mu$s to $\sim$6.5 $\mu$s.  
The random number generator generates pseudo-random numbers with a uniform distribution.
These uniformly distributed random numbers, and the Gaussian-distributed ones used to generate the amplitude modulated noise and additive noise, were produced through
the Mersenne twister algorithm \citep{mn98} which is known to have a low auto-correlation \citep[see, e.g.,][]{h13} and produces $2^{19937}-1$ random results before repeating.
We use the same seed at the start of each run but vary the seed for different trials so that the input scattering timescales do not change every time the code is run.

\begin{figure*}
    \centering
\includegraphics[width=4.5in]{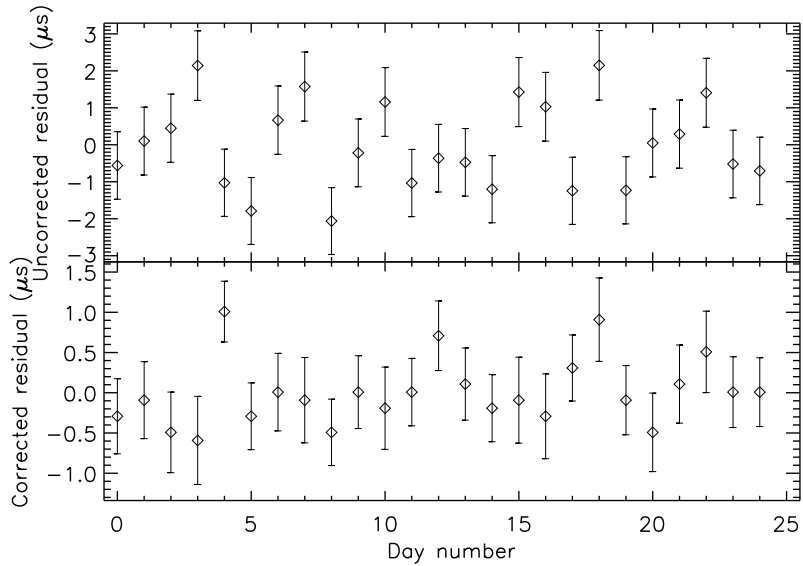}
        \label{fig:delays1}
\caption{Top: Uncorrected residuals for 25 trials where the input scattering times were drawn from a random distribution with a mean of 5 $\mu$s and a standard deviation of 1 $\mu$s. 
%The additive noise level was set to 0.5 times the average on-pulse power for a single pulse and $10^4$ pulses have been added to form the average profile of each trial.
Bottom: Delays after applying the CS correction. The RMS of uncorrected and CS-corrected residuals are 1031 ns and 407 ns, respectively.}
\label{fig:delays_WDS}
\end{figure*}

In order to gauge the effect of the scattering correction on timing measurements, we first calculated the TOAs for the scatter-broadened pulse profiles on the 25 trial epochs using the Taylor algorithm \citep{t92}.
This cross-correlation algorithm calculates the relative time delay offset between the observed average profile and a high quality standard profile through an iterative, frequency-domain fitting algorithm to calculate a TOA.
A scattered standard profile which has a scattering timescale equal to the mean of the input scattering times (5$\mu$s) was used for this purpose.
This is a realistic assumption for actual PTA observations, as standard profiles are formed from averages of scattered observed profiles.
The top plot of Figure~\ref{fig:delays_WDS} shows the residuals before scatter correction for 25 trials.     
The RMS of the residuals derived from the Taylor algorithm analysis for scatter-broadened average profiles have a standard deviation of 1~$\mu$s over the 25 trials of observation.

The IRFs were recovered from the procedure described in the beginning of Section~3.
They were 
 smoothed by eight samples, and a one-sided exponential function of the form $h_I(t)=|A U(t-t_0)\,{\rm exp}^{-(t-t_0)/\tau^{\prime}}|^2$, where $A$ is the amplitude and other terms are as explained in Section~3, was fit to the smoothed IRFs.
The best fitting parameters for amplitude $A$, time delay $t_0$ and scattering timescale $\tau$ of the recovered IRFs were determined by minimizing a $\chi^2$ grid search over the three parameters.
As shown in Figure~\ref{fig:PBF_WDS}, the intensity IRF with a scattering timescale of 5.1 $\mu$s input to the simulation is recovered with a scattering timescale of 5.7 $\pm$ 0.4 $\mu$s.
%The PBFs were recovered from cyclic spectra on each of these trials.
In a production timing campaign, the IRFs reconstructed here would need to be analyzed further in a multi-frequency model from which terms with different frequency scalings -- including the important frequency-independent term -- would be extracted.

As previously described, we find the best fitting $t_0$ values or the scatter-corrected TOAs.
The residuals were calculated by subtracting these scatter corrected TOAs from those predicted by a simple timing model.
The errors on the best fit parameters $t_0$ and $\tau$ were calculated from the $2\times2$ covariance matrix
evaluated from the second derivatives of the $\chi^2$.% at the solution point.
The error on the residuals is the error of $t_0$.
When scatter corrected, the RMS of the residuals reduces to 407 ns from 1 $\mu$s.
These results show that the RMS of the residuals calculated from scatter-corrected TOAs is significantly lower than  the RMS of the residuals of scattered profiles calculated using the TOAs from the Taylor algorithm, approaching a factor of 2.5 improvement.% in optimal cases.

In the absence of other effects such as pulse-to-pulse jitter and red noise, TOA errors are dominated by radiometer noise that scales as $W/(S_{0}\sqrt N)$. 
Here $W$ is the width of the pulse profile, $S_{0}$ is the  single pulse signal-to-noise-ratio, and $N$ is the number of pulses averaged.
In order to test the correction scheme for various S/N levels, defined as the ratio between the amplitude of the pulse peak and the RMS of the profile noise baseline, 
the 
level of additive noise, or the amount of instrumental noise in the cyclic spectrum, was varied, 
which resulted in pulse profiles with S/N ranging from approximately 100 to 2500.
We have used 128 profile bins.

\begin{figure*}
    \centering
\includegraphics[width=4.5in]{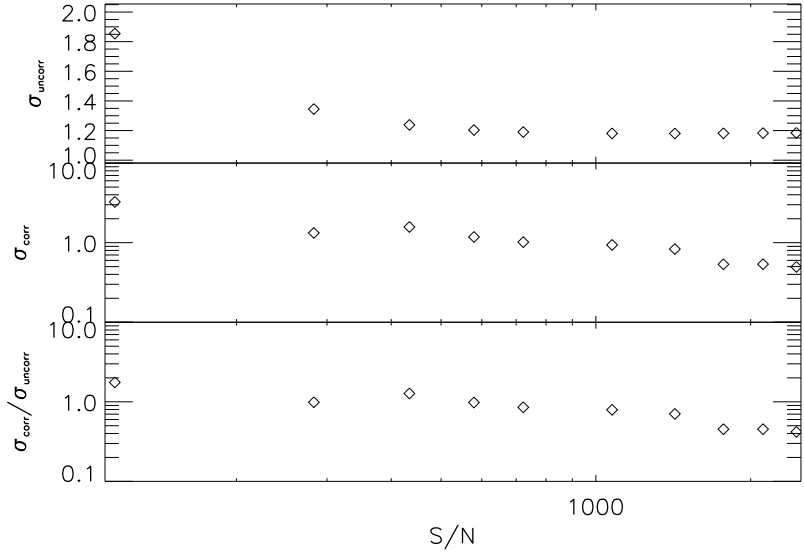}
        \label{fig:snr}
\caption{RMS of uncorrected residuals (top), RMS of CS-corrected residuals (middle), 
and the ratio between the RMS of CS-corrected residuals and the RMS of uncorrected residuals (bottom) as a function of average profile S/N. 
The RMS of uncorrected residuals ranges from 1.8 at S/N 116 to 1.1 at S/N 2450. 
Each RMS value is calculated over 25 trials where the scattering timescale of each average profile was drawn from a uniform distribution with a mean of 5 $\mu$s and a standard deviation of 1 $\mu$s. 
The S/N was varied by changing the amount of additive noise. 
}
\label{fig:rms_vary_snr_WDS}
\end{figure*}

\begin{figure*}
    \centering
\includegraphics[width=4.8in]{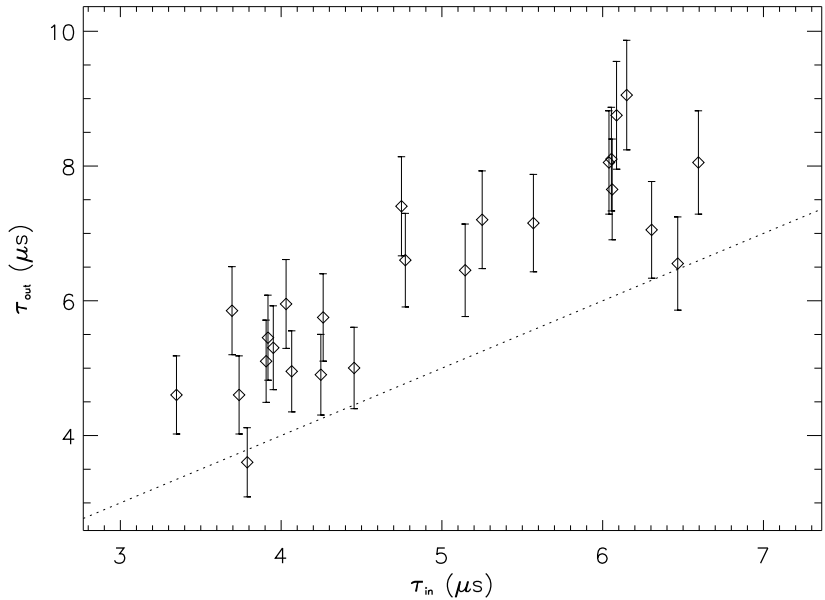}
        \label{fig:four_average}
\caption{ 
The scattering timescales from the best fitting one-sided exponential functions vs input scattering timescales for S/N value 2500, corresponding to the highest S/N data point in Figure~\ref{fig:rms_vary_snr_WDS}. 
The error bars are calculated from the covariance matrix from the one-sided exponential fits.
The dashed line marks the $y=x$ line. 
The fitting procedure tends to over-estimate the scattering timescale by roughly $30\%$ in most cases.
Simulations indicate that smoothing is likely responsible for over-estimating the scattering timescales by 10\%. When not smoothed, the timescales are underestimated by roughly the same fraction.
However, the lowest RMS residuals are obtained when the fits are performed on smoothed IRFs.
}
\label{fig:tau_SNR_WDS}
\end{figure*}

Figure~\ref{fig:rms_vary_snr_WDS} shows that the RMS of scatter-corrected residuals decreases with increasing S/N.
We find that the shape of the recovered IRFs for low S/N profiles whose S/N is lower than $\sim$100 
%or when the number of pulses averaged over is less than 100, 
starts to differ significantly from the input IRF. 
For typical NANOGrav observations, more than 500 pulses are averaged per TOA and resultant S/N values are typically greater than 100 (though not for all cases).
We note that the IRF does not change appreciably from ISM variations on the timescale required to accumulate a S/N of greater than 100.
In Figure~\ref{fig:tau_SNR_WDS} we show the input scattering timescales and the scattering timescales recovered from the best fitting one-sided exponential functions.

\begin{figure*}
    \centering
\includegraphics[width=4.5in]{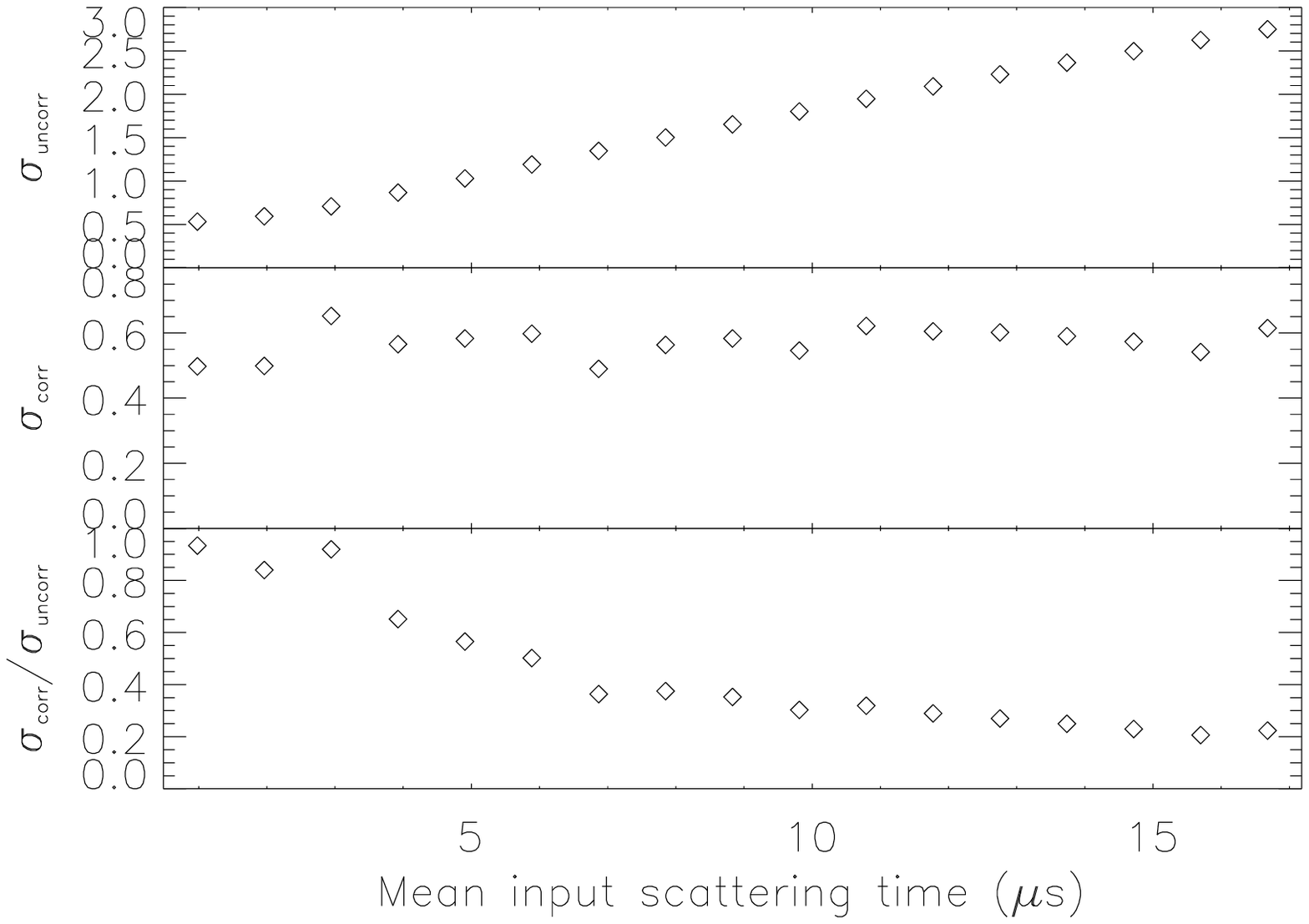}
        \label{fig:rams_ratio}
\caption{RMS of uncorrected residuals (top), RMS of CS-corrected residuals (middle), and the ratio between the RMS of CS-corrected residuals and the RMS of uncorrected residuals (bottom).
The scattering times for each mean scattering time are drawn from a random distribution that has the given mean and a standard deviation equal to $\rm mean/5$. Therefore, as the mean scattering time increases, the variation of the scattering time also increases as is the case in real pulsars. 
The profile S/N used is $\sim$ 8000.
}
\label{fig:rms_vary_meantau_WDS}
\end{figure*}

We have also assessed the effectiveness of the CS scatter correction for varying mean scattering times ranging from 1 to 17~$\mu$s.
Figure~\ref{fig:rms_vary_meantau_WDS} shows the RMS of residuals before scatter correction, RMS of residuals after scatter correction, and the ratio between the corrected and uncorrected RMS. 
For each value of the mean scattering time, the RMS of the input scattering times was set to be a factor of 0.2 of the mean scattering time, so that the RMS of scattering times increase with increasing mean scattering time, as typically seen in real pulsar signals \citep{hs08,rdb06}.
We find that the ratio between the CS corrected and uncorrected RMS residuals decreases with increasing mean scattering timescale.
Profiles with S/N $\sim$ 8000 were used in this simulation.
The scatter correction process improves for longer scattering timescales due to the possibility of obtaining better fits to the recovered IRFs.
These results are expected since even in an ideal situation the scatter correction scheme will be limited by the finite S/N of the simulations.

\subsection{Presence of non-scattering delays}

As emphasized previously, $t_0$ as defined in Equation~\ref{eq:ht} may include other chromatic and achromatic delays.
We do not address these chromatic effects, in general, here; however, our single-frequency results could be generalized to a more realistic multi-frequency correction procedure.
In order to verify that the scatter correction process does not remove other delays present, 
we added a sinusoid to the data (e.g. a systematic effect or a GW)
%we added a GW signal 
to the simulated data and performed the scatter correction.
The simulated sinusoidal signal has a period of 25 days and an amplitude of 0.5 $\mu$s.
The sinusoid was sampled at 25 trial epochs to obtain the delay values which were added as a time shift, to the scatter broadened simulated pulse profiles.
The scattering timescales of these profiles were drawn from a random distribution with a variance of 1 $\mu$s, as outlined in Section 3.1. 
The corresponding IRFs were then recovered from the pulse profiles using the CS method.
These profiles have been averaged over 10,000 single pulses and have a S/N of $\sim$ 1000.
These IRFs were smoothed by eight samples, and a one-sided exponential function was fit as explained in Section~3.1 to each recovered IRF.
 
\begin{figure*}
    \centering
{
        \includegraphics[width=5.2in]{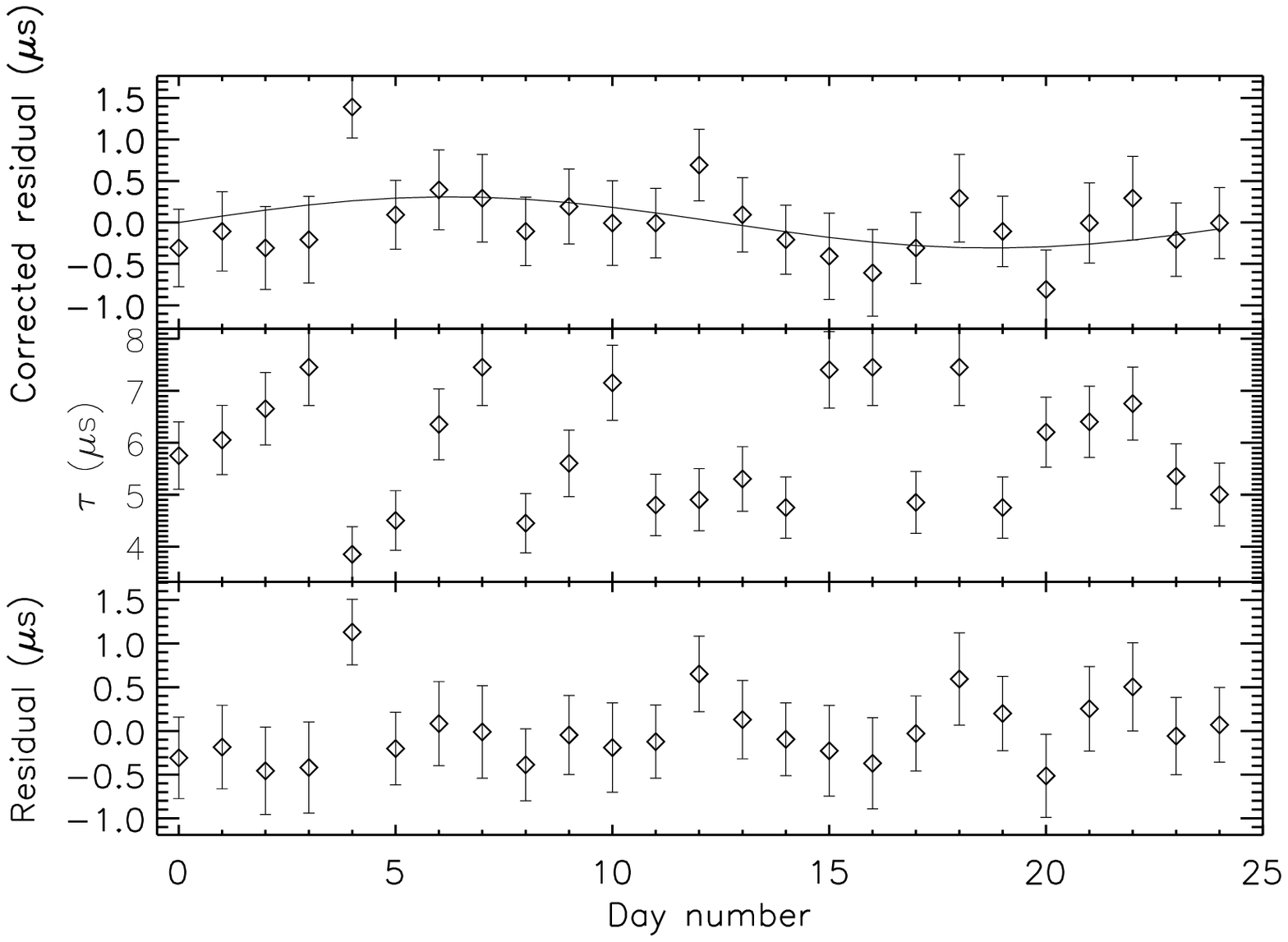}
        \label{fig:jitter_profs}
    }
\caption{Scatter corrected residuals (top), $\tau$ (middle) for a function of the form $h(t)=A U(t-t_0)\,{\rm exp}^{-(t-t_0)/\tau^{\prime}}$, where $\tau=\tau^{\prime}/2$, and the white residuals which are calculated by removing the sinusoid from the scatter corrected residuals.
The best fit sinusoid, which represents a gravitational wave signal, is overplotted in the top plot.
The recovered sinusoid has a period of 25.022 $\pm$ 0.005 days and an amplitude of 0.3087 $\pm$ 0.0004 $\mu$s. 
The RMS of the white residuals is 390 ns. The RMS of the input scattering timescales for this case is $\sim$ 1 $\mu$s. 
}
\label{fig:t0_tau_gw_WDS}
\end{figure*}

In order to compare with the previous results of scatter-corrected residuals, we fit a sinusoid to the best fit $t_0$ values and subtract it off, in order to get the expected white residuals.
We fit for the period and amplitude of the sinusoid and find the best fitting values to be 25.022 $\pm$ 0.005 days for period and 0.3087 $\pm$ 0.0004 $\mu$s for amplitude. 
The RMS of residuals after the sinusoid was fit out is 390 ns.
These results are illustrated in Figure~\ref{fig:t0_tau_gw_WDS}, which shows the best fitting parameters $t_0$ and $\tau$, and the residuals when the sinusoid is removed from the best fit $t_0$ values. 
The errors on the best fit parameters $t_0$, $\tau$ and the best fitting sinusoidal signal, were calculated from the $2\times2$ covariance matrix evaluated from the second derivatives of the $\chi^2$.
The error on the residuals is the error of $t_0$. 

Through an alternate method, frequency-dependent (chromatic)
delays from achromatic delays such as those due to a gravitational wave can be separated though 
a carefully-designed multi-frequency timing analysis.
What is gained with the CS-approach is that we retrieve a {\em function} for the IRF from which we can extract a TOA. 
There could be multiple ways in which we could do this. 
In the case of a simple one-sided exponential function we can consider the rising edge of the recovered IRF as the scatter corrected TOA. 
A better algorithm, and one which would account for other functional forms, would be to calculate the scattering delay by fitting for the frequency-dependent and independent parts of the IRF.  
But, a full implementation of that program requires a careful analysis of the form
of the IRF due to various ISM propagation effects and is beyond the scope of
this paper.
Nevertheless, the additional information provided by the IRF function, which is itself
frequency-dependent, should allow this separation to be made much more accurately
than when only frequency-dependent TOAs (single numbers) are produced.

\subsection{Thick screen results}

As mentioned earlier, the one-sided exponential shape results from assuming that the ISM acts as a transverse thin screen situated in between the pulsar and the observer.
IRFs for realistic pulsar signals being scattered off the ionized ISM with electron density fluctuations of various length scales and different levels of turbulence may deviate from this simple form.
\citet{w72, w73} developed analytical solutions to various situations such as scattering from a thick screen situated either near the observer or pulsar or in between the two, or two thin screens and a uniform medium extending from the pulsar to the observer.
The main characteristics of these functions are a slower rise time and an exponential decay.
For example,
for a thick screen the IRF \citep{w73} is given by 
\begin{equation}
h(t)=n_2(t) \sqrt\frac{\pi\tau^{\prime}}{4 t^3}\,{\rm exp}\left[-\frac{\pi^2\tau^{\prime}}{16 t}\right].
\label{eq:thick_screen_eq}
\end{equation}
where $n_2(t)$ is complex Gaussian white noise and $\tau^{\prime}$ is the scattering timescale in the voltage IRF.
This function has a slower rise time 
and the peak of the scattering broadened pulse will be at 0.41$\tau^{\prime}$ \citep{w72} where $\tau$ is the scattering timescale.

\begin{figure}[h]
\begin{center}
\includegraphics[width=3.5in]{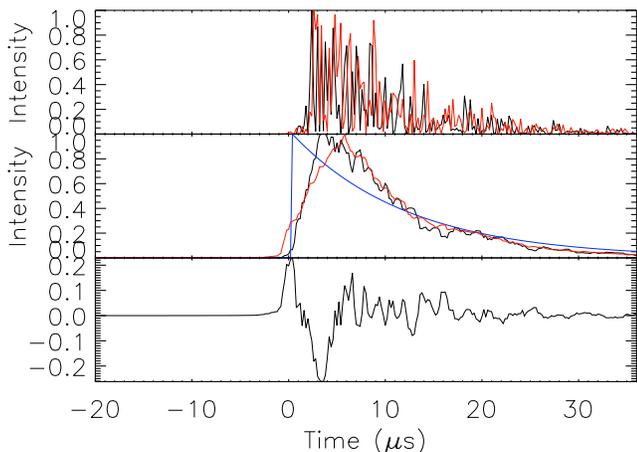}
\caption{An example recovered normalized intensity IRF $h_{I}(t)$ for a thick screen given by Equation~\ref{eq:thick_screen_eq} constructed from a cyclic spectrum (red curve)
for an input scattering timescale of 5 $\mu$s.
For this realization, the scattering timescale estimated using the fitting technique is 10.6 $\mu$s.
The input function is shown by the black curve.
the best fitting one-sided exponential function is shown by the blue curve. 
The middle plots are smoothed versions of the top plots and 
the bottom plot shows the residual between the input and recovered IRFs.
}
\label{fig:PBF_thick_WDS}
\end{center}
\end{figure}

\begin{figure*}
    \centering
\includegraphics[width=4.5in]{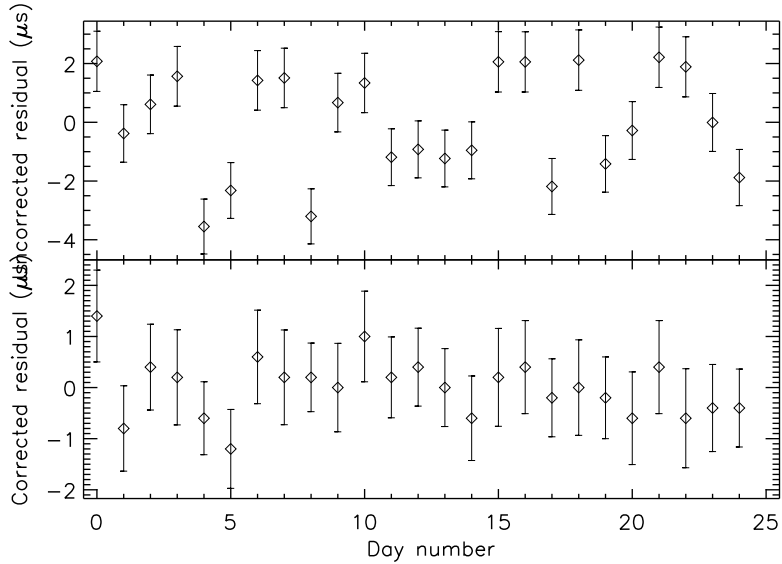}
\caption{Top: Uncorrected residuals for 25 trials where the input scattering times were drawn from a random distribution with a mean of 5 $\mu$s and a standard deviation of 1 $\mu$s for a thick screen. 
$10^4$ pulses have been added to form the average profile of each trial.
Bottom: Delays after applying the CS correction. The RMS of uncorrected and CS-corrected residuals are 1890 ns and 580 ns, respectively.}
\label{fig:delays_thick_WDS}
\end{figure*}

Figure~\ref{fig:PBF_thick_WDS} shows example input and recovered IRFs along with the best fitting one-sided exponential for a scattering timescale of 5 $\mu$s.
We note that when the IRF envelope is modulated by noise due to scintillation, the IRF of a thick screen deviates very little from the one-sided exponential IRF of a thin screen.
Therefore, we fit one-sided exponential functions for the recovered IRFs in order to extract the scatter corrected TOA $t_0$ and the scattering timescale $\tau$.
Figure~\ref{fig:delays_thick_WDS} we show the uncorrected and scatter corrected TOAs over 25 days and a mean scattering timescale of 5 $\mu$s in the thick screen scenario.
For profiles containing a mean scattering delay of 5 $\mu$s and an RMS variation of $\sim$1~$\mu$s, the RMS of uncorrected residuals is $\sim$1.8~$\mu$s and CS-correction technique reduced the RMS variation to $\sim$580 ns.

\subsection{Effects of pulse-to-pulse jitter}

Single pulses from some millisecond pulsars show random fluctuations in arrival phase which can be of the order of a pulse width.
This is referred to as pulse phase jitter \citep{cs10}.
While pulses from PSR B1937$+$21 show jitter that contributes to a TOA uncertainty of 19 ns \citep{cs10}, giant pulses from this object also exhibit jitter \citep{kt00}.
PSRs J1713$+$0747 and J0437$-$4715, precisely timed PTA pulsars (epoch averaged RMS of 30 ns \citep{dfg12} and 75 ns \citep{mhb12}), are among the MSPs that show significant phase jitter that affect the TOA errors at levels of 65 ns and 105 ns \citep{cs10} respectively.
In order to test the effect of phase jitter on the CS recovery of IRFs, we simulated pulsar emission with jitter where the pulse phase fluctuations were drawn from a Gaussian random distribution, consistent with observations of jitter in normal pulsars.

We have conducted this analysis for four values of the jitter parameter $F_{J}$, which is defined as $F_{J}^2=1-\left(W_1^2/W_a^2\right)$ \citep{cs10}, where $W_1$ and $W_a$ are the widths of single pulses and average profile respectively.
Single pulses with widths 40, 30, 20, and 10 $\mu$s and a scattering timescale of 5 $\mu$s were shifted in pulse phase from the center of the pulse to form a jittered average profile of width $\sim$40 $\mu$s.
These correspond to jitter parameter values of 0.00, 0.66, 0.87 and 0.97.
We have also computed the effect of scatter correction on jittered average profiles on 25 epochs for the above four jitter parameter values.
We find that the RMS correction ratio ($\sigma_{corr}/\sigma_{uncorr}$) increases from  0.3 to 1.3 as the jitter parameter increases from 0.0 to 0.97.
Given the fact that real MSP signals show small amounts of jitter \citep{cs10},
the effect of phase jitter on impulse response recovery should be minimal.

\section{Discussion}
We have used simulations to show that the CS method, based on electric field phase information, can be used to 
 accurately recover the complex voltage IRF, $\it h(t)$, under realizable conditions for the brightest pulsars observed with 100-m class radio telescopes, for which the S/N we have used here is applicable.
For profiles containing a mean scattering delay of 5 $\mu$s and an RMS variation of $\sim$1~$\mu$s, the CS-correction technique reduced the RMS variation to 407 ns.
This is a factor of $\sim$2.5 decrease in the residual RMS.
We also find that the ratio of pre to post scatter-correction RMS improves with increasing profile S/N. This implies that CS scatter correction is, not surprisingly, more effective on bright pulsars.
The finite S/N of the simulation sets a lower limit on the efficacy of the CS correction scheme.
In the presence of a GW signal, in order to separate the scattering delays from the GW delays, it becomes necessary to fit a one-sided exponential function to the recovered IRF, in order to locate the rising edge of the function, which gives the scatter corrected TOA.
Real pulsar signals will contain other delays, such as those due to refraction, that will cause $t_0$ of Equation~\ref{eq:ht} to be chromatic.

Out of the  MSPs that exhibit jitter, jitter parameters of $\sim$ 0.2--0.5 for giant pulses of B1937$+$21 \citep{kt00}, 0.4 for pulses of J1713$+0$747 \citep{sc12}, and 0.07 for pulses of J0437$-$4715 \citep{lkl11} have been observed.
Our results show that one-sided exponential IRFs with the expected timescales can be recovered from pulse profiles that show these amounts of jitter using this technique.
Therefore the effect of pulse phase jitter on IRF recovery should be negligible.

It is important to note that the improvement in timing precision that we demonstrate is solely due to scattering time delay correction and not to pulse sharpening through the removal
of scatter broadening.
The advantage of this method is that it does not require prior assumptions of the shape of the IRF or the pulse shape, unlike the previously proposed IRF retrieval techniques that rely on prior assumptions of either one or both entities.
The strong frequency dependence of scattering can  help distinguish between ISM effects and other achromatic effects when multi-frequency observations are made.
This is vital because one of the key applications for this technique is in the effort
to detect GWs with pulsar timing.
Since the GW signal is achromatic, we must be careful to prevent or minimize its inclusion
in any correction of TOAs that is done in a dedicated PTA effort.

This problem is not limited to the IRF estimation technique presented here.
It is inherent in separating the effects of ISM and other frequency-dependent delays
from the achromatic signal of interest. 
Following the pioneering work of \citet{d11} and \citet{wdv13},
our simulation takes a next step toward developing a production-quality
chromatic correction technique.
As we have emphasized, the CS-technique produces a IRF {\em function} from which we can extract a TOA.
This provides the platform for a  fuller multi-frequency analysis of the
timing information embodied in $h(t; \nu)$ which should certainly
improve our ability to separate chromatic and achromatic influences
on the pulse arrival time.

The simulated pulsar signal is scatter-broadened amplitude-modulated noise and includes scintillation effects.
We have limited the analysis to this case in order to demonstrate the effect of scatter correction on improving timing residuals without the complications of other smaller effects.
This technique can be applied to real pulsar signals whose phase information is preserved when recorded with a baseband setup.
However, when applying this to real pulsar data we note that the scintle size, phase connection between scintles, and signal-to-noise ratio may need to be accounted for.
Furthermore, when implementing on real pulsar data, the cyclic spectra calculated at radio frequencies $\nu$ within a bandwidth of $B$ will be valid within a region described by $|\alpha /2|+|\nu|< B/2$ \citep[see][for details]{d11}.
Cyclic spectroscopy has so far been tested only on pulsar B1937$+$21 \citep{d11}, a bright pulsar which exhibits pronounced DM variations and pulse broadening variations \citep{rdb06}; the possibility of achieving a descattered pulse profile is shown in that paper. 

Scattering correction, when fully achieved, will allow higher precision pulsar timing, which will facilitate GW detection efforts using pulsars. It will also increase the number of MSPs able to be included in a pulsar timing array and will improve timing at lower frequencies.
This should also improve timing of pulsars if found in the Galactic center region, currently limited due to scattering effects from turbulent plasma in these dense regions.

The authors thank the referee Mark Walker, whose suggestions have improved this work. The authors also thank the NANOGrav collaboration and M. Lam, J. Lazio, and T. Pennucci for a critical reading of the manuscript.

\end{document}